\documentclass[journal,twocolumn]{IEEEtran}
\usepackage{amsfonts}

\usepackage{bbm}
\usepackage{enumerate}
\usepackage{amsmath,amsthm}
\usepackage{mathtools}
\usepackage{algorithm,algorithmic}
\usepackage{float}
\usepackage{hyperref}
\usepackage{color}
\usepackage{makeidx}
\usepackage{bbm}
\usepackage{graphicx}
\usepackage{lipsum}
\usepackage{soul}
\usepackage{tabularx}
\usepackage{dsfont}
\usepackage[table,xcdraw]{xcolor}

\usepackage{amsfonts}
\usepackage{times}
\usepackage{graphicx}
\usepackage{latexsym}
\usepackage{dsfont}
\usepackage{amssymb}
\usepackage{amsmath}
\usepackage{cite}
\usepackage{verbatim}
\usepackage{subfigure}

\def\bb0{{\mathbb{0}}}

\def\bb{{\mathbf{b}}}

\def\bh{{\mathbf{h}}}

\def\b0{{\mathbf{0}}}

\def\sf0{{\mathsf{0}}}

\usepackage{epstopdf}

\newcommand{\sref}[1]{{Section}~\ref{#1}}
\newcommand{\fref}[1]{{Fig.}~\ref{#1}}

\newcommand{\subto}{\operatorname{s.t.}}

\begin{document}
\title{ Learning Reflection Beamforming Codebooks for  \\ Arbitrary RIS and Non-Stationary Channels}
\author{Arizona State University}
\author{Yu Zhang and Ahmed Alkhateeb \thanks{Yu Zhang and Ahmed Alkhateeb are with Arizona State University (Email: y.zhang, alkhateeb@asu.edu). This work is supported by the National Science Foundation under Grant No. 1923676.}}
\maketitle

\begin{abstract}
Reconfigurable intelligent surfaces (RIS) are expected to play an important role in future wireless communication systems. These surfaces typically rely on their reflection beamforming codebooks to reflect and focus the signal on the target receivers. Prior work has mainly considered pre-defined RIS beamsteering codebooks that do not adapt to the environment and hardware and lead to large beam training overhead. In this work, a novel deep reinforcement learning based framework is developed to efficiently construct the RIS reflection beam codebook. This framework adopts a multi-level design approach that transfers the learning between the multiple RIS subarrays, which speeds up the learning convergence and highly reduces the computational complexity for  extremely large RIS surfaces. The proposed approach is generic for co-located/distributed RIS surfaces with arbitrary array geometries and with stationary/non-stationary channels. Further, the developed solution does not require explicit channel knowledge and adapts the codebook beams to the surrounding environment, user distribution, and hardware characteristics. Simulation results show that the proposed learning framework can learn optimized interaction codebooks within reasonable iterations. For example, with only 6 beams, the learned beam codebook outperforms a 256-beam DFT codebook, which significantly reduces the RIS beam training overhead.
\end{abstract}

\section{Introduction} \label{sec:Intro}

Reconfigurable intelligent surfaces (RIS) are envisioned as a key enabler in extending coverage and overcoming blockage in millimeter wave (mmWave) and terahertz (THz) communication systems \cite{Taha2021}. Realizing the potential gains of these surfaces, however, relies on carefully designing reflection beamforming vectors that reflect the incident signals and focus them on the target receivers. This is normally done by pre-designing a reflection beamforming codebook that can scan all the directions, such as DFT codebooks. This, however, is associated with several challenges in extremely large reconfigurable intelligent surfaces: (i) The beamforming codebook design typically relies on channel knowledge which is very hard to acquire in large RIS systems with nearly-passive elements, (ii) the pre-defined codebooks  normally requires huge beam training overhead, and are not adaptive to the site-specific environment, user distribution, and hardware characteristics, (iii) the surfaces may have different visibility regions leading to non-stationary channels across the surfaces \cite{Carvalho2019}, and (iv) these surfaces could generally be distributed or could have arbitrary array geometries that are hard to model.  This motivates the development of novel approaches for the design of the RIS reflection/beamforming codebooks.

\textbf{Contribution:} In this paper, we develop a low-complexity yet efficient deep reinforcement learning (DRL) approach for designing RIS reflection beam codebooks. The proposed solution has several advantages: (i) It can be applied to arbitrary (centralized or distributed) RIS surfaces with unknown array geometry, (ii) it accounts for both stationary and non-stationary channel models, (iii) the design of the DRL model accounts for the practical RIS hardware limitations, such as the quantized phase shifter constraints on the RIS elements, (iv) the proposed approach does not require any explicit channel knowledge and relies only on receive power measurements, which relaxes the synchronization requirements and the channel estimation overhead. Further, our developed solution includes a novel successive learning and combining framework that highly reduces the convergence time, which is crucial for lowering the computational complexity of the large RIS surfaces. The simulation results highlight the capability of the proposed solution in efficiently learning codebooks that adapt to user distributions and channel characteristics (such as non-stationarity). Besides, the learned codebooks outperform DFT codebook with much smaller codebook size, which significantly reduces the beam training overhead.

\textbf{Prior Work:} The existing work for RIS reflection pattern design normally requires explicit channel knowledge \cite{Yan2020}, and does not focus on designing a codebook \cite{Liu2021}. Furthermore, the reflecting elements are normally assumed to have continuous phases for the ease of optimization \cite{Rehman2021}. Besides, the prior work on  RIS beamforming design \cite{Yan2020,Liu2021,Rehman2021} generally ignored the possible non-stationarity of the RIS channels \cite{Carvalho2019}.

\section{System and Channel Models} \label{sec:System}

The codebook design approach developed in this paper can be applied to various RIS deployments including scenarios with distributed RIS sub-surfaces and with stationary and non-stationary channel models. Therefore, we adopt generic system and channel models, as described in the next subsections.

\subsection{System Model}

\begin{figure}[t]
	\centering
	\includegraphics[width=.95\columnwidth]{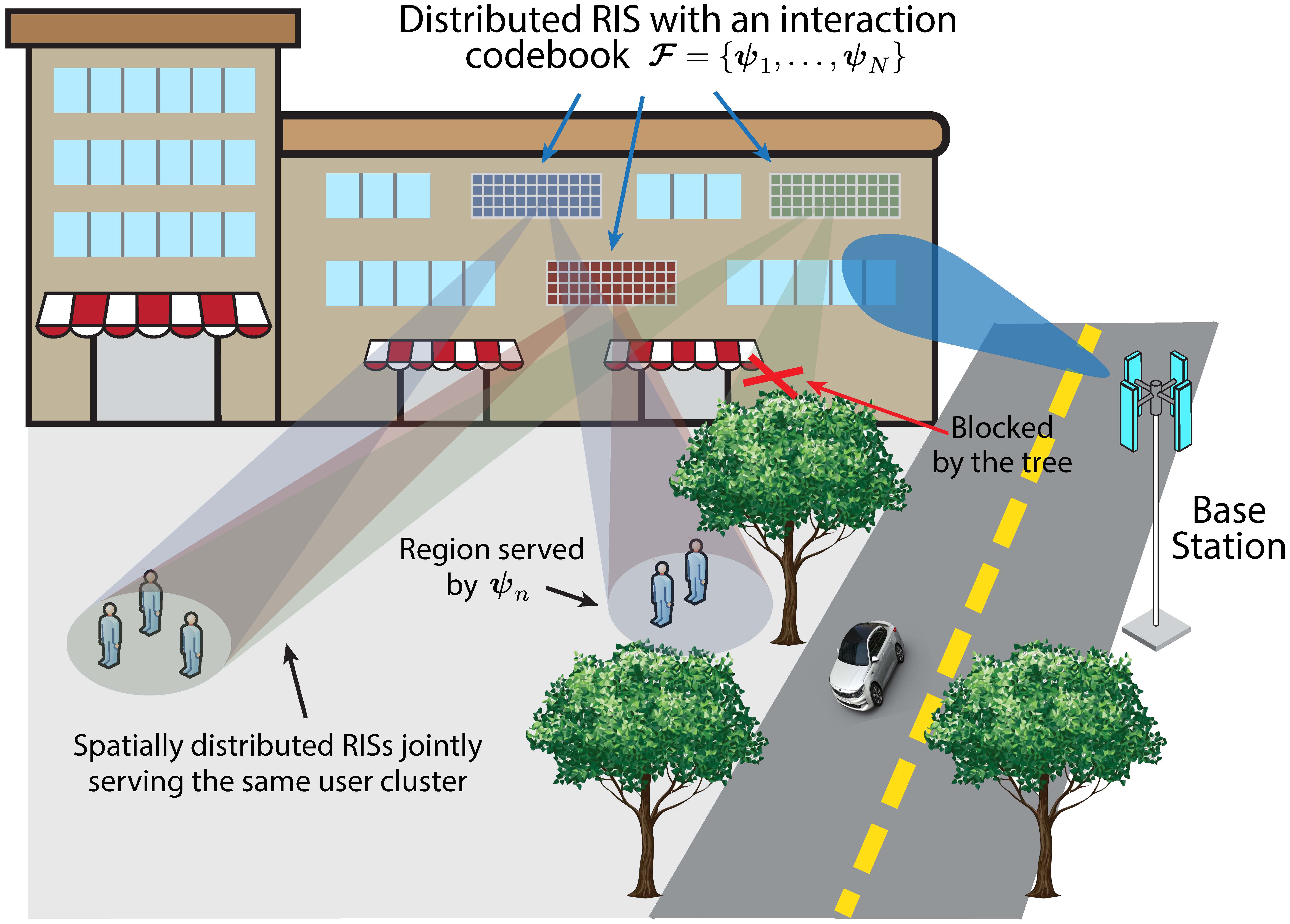}
    \caption{This figure illustrates the adopted system model where a mmWave/THz base station is communicating with multiple users via a RIS. An interaction codebook is used in the RIS in order to serve users in different clusters.}
	\label{fig:sys}
\end{figure}

We consider the system where a base station, acting as a transmitter, is communicating with a receiver through a RIS. The RIS has $M$ interaction (reflection) elements, which can be generally distributed over multiple sub-surfaces as shown in \fref{fig:sys}. For simplicity, both the transmitter and receiver are assumed to have single antenna. The considered scenario is assumed to have no direct link between the transmitter and receiver. This models the situation where the direct link is either blocked or has negligible receive power compared to that received through the RIS-assisted link.
With these assumptions, if the transmitter sends a symbol $s_u \in \mathbb{C}$ to the $u$-th receiver, and the RIS uses an interaction vector ${\boldsymbol\psi}$ to reflect the impinging signals, the received signal at the $u$-th receiver can be expressed as
\begin{equation}\label{rec}
  y_u = {\bh}_{R, u}^T{\boldsymbol\Psi}{\bh}_Ts_u + n_u,
\end{equation}
where ${\bh}_T$ and ${\bh}_{R, u}$ denote the channels between transmitter and RIS, and between RIS and the $u$-th receiver's antenna, respectively. The transmitted symbol satisfies the average power constraint $\mathbb{E}[|s_u|^2]=P_s$ and $n_u\sim\mathcal{N}_\mathbb{C}(0, \sigma_n^2)$ is the receive noise at the receiver. The $M\times M$ diagonal matrix ${\boldsymbol\Psi}$ is the interaction matrix of the RIS, i.e., ${\boldsymbol\Psi} = \text{diag}({\boldsymbol\psi})$. Therefore, we can rewrite \eqref{rec} as
\begin{equation}\label{re-rec}
  y_u = ({\bh}_T\odot{\bh}_{R, u})^T{\boldsymbol\psi}s_u + n_u,
\end{equation}
where $\odot$ is the Hadamard product.

To reduce the precoding optimization complexity, these large surfaces typically adopt interaction (reflection) beam codebooks. Let $\boldsymbol{\mathcal{F}}$ denote the interaction codebook that contains $N$ interaction vectors, with the $n$-th interaction vector given by
\begin{equation}\label{diag}
  {\boldsymbol\psi}_n = \frac{1}{\sqrt{M}}[e^{j\theta_{1n}}, \dots, e^{j\theta_{Mn}}]^T, ~ \forall n=1,2,\dots,N.
\end{equation}
Besides, to account for the practical hardware limitations on the RIS surfaces, we assume that the RIS can only adjust the phases of the incident signals at each element with limited resolution, via $q$-bit quantized phase shifters, while keeping their magnitudes unchanged. Therefore, each phase shift $\theta_{mn}$ in \eqref{diag} is selected from a finite set $\boldsymbol\Theta$ with $2^q$ possible discrete values drawn uniformly from $(-\pi, \pi]$.

\subsection{Channel Model}

We consider two different channel models in this paper, namely, stationary and non-stationary channel models. We will use the channel between the transmitter and the RIS as an example and the channels between the receivers and the RIS can be defined in a similar fashion.

\textbf{Non-stationary channel model:}
Due to the large physical size of the RIS, non-stationarity could be observed across the elements on the surface. This means that each element in RIS has its own set of observable users/clusters, number of multi-paths, and seeable arrival/departure angle ranges etc. Equivalently, each user/cluster can only see a subset of reflecting elements out of the whole surface, which is defined as visibility regions \cite{Carvalho2019}.
Therefore, we adopt a channel model that takes the non-stationarity factor into account. Specifically, we assume that the channel between the transmitter and the $m$-th interaction element of RIS can be expressed as
\begin{equation}\label{non-stat-ch}
[\bh]_{m} = \sum_{\ell=1}^{L} \mathbbm{1}_m\{\phi_{\ell}\} \alpha_{\ell} \mathbf{a}\left(\phi_{\ell}\right), ~ \forall m=1,\dots,M,
\end{equation}
where $L$ is the total number of multi-paths, $\alpha_{\ell}$ and $\phi_{\ell}$ are the complex gain and angle of arrival (AoA) of the $\ell$-th path. $\mathbf{a}\left(\phi_{\ell}\right)$ is the array response vector of the RIS, the definition of which depends on the adopted surface geometry.
$\mathbbm{1}_m\{\cdot\}$ is the indicator function, which is defined as
\begin{equation}\label{indic}
\mathbbm{1}_m\{\phi_{\ell}\} = \left\{
\begin{array}{ll}
1, & \text{if $\phi_{\ell}\in [\phi_{\min}^{(m)}, \phi_{\max}^{(m)}]$}, \\
0, & \text{otherwise},
\end{array}
\right.
\end{equation}
with $[\phi_{\min}^{(m)}, \phi_{\max}^{(m)}]$ representing the range of seeable AoA of the $m$-th RIS element.
The \textbf{stationary channel model} can be viewed as a special case of non-stationary channel model defined above by removing the indicator function.

\section{Problem Formulation} \label{sec:Prob}

In this paper, we investigate the development of an RIS reflection codebook design approach that adapts to the RIS geometry, channel model, user distribution, and hardware impairments. Next, we formulate the key codebook design problem before addressing it in \sref{sec:Solution}.
Given the system and channel models described in \sref{sec:System}, if the $u$-th receiver is served by the $n$-th interaction vector $\boldsymbol{\psi}_n\in\boldsymbol{\mathcal{F}}$, then its composite channel gain is given by
\begin{align}\label{sig-power}
  g_{u, n} = & \left|({\bh}_T\odot{\bh}_{R, u})^T{\boldsymbol{\psi}_n}\right|^2, \\
  = & \frac{1}{M}\left| \sum_{m=1}^M \alpha_{T, m}\alpha_{R, u, m}e^{j(\phi_{T, m}+\phi_{R, u, m}+\theta_{mn})} \right|^2, \label{complex-sum}
\end{align}
where $\theta_{mn}=\angle[\boldsymbol{\psi}_n]_m$ is the phase of the $m$-th element of the $n$-th interaction vector,
$\alpha_{T, m}$ and $\phi_{T, m}$ are the gain and the phase of the channel between the transmitter and the $m$-th element of the RIS,
$\alpha_{R, u, m}$ and $\phi_{R, u, m}$ are the gain and the phase of the channel between the $u$-th receiver and the $m$-th element of the RIS.
The RIS reflection codebook design problem can then be formulated as
\begin{align}\label{Prob}
  \boldsymbol{\mathcal{F}}_{\mathsf{opt}}=\mathop{\arg\max}_{\boldsymbol{\mathcal{F}}} & \hspace{10pt} \frac{1}{|\boldsymbol{\mathcal{H}}_R|}\sum_{{\bh}_{R, u}\in{\boldsymbol{\mathcal{H}}_R}} \max_{\boldsymbol{\psi}_n\in\boldsymbol{\mathcal{F}}} g_{u, n} \\
  \subto & \hspace{10pt} |\boldsymbol{\mathcal{F}}| = N, \\
  & \hspace{10pt} \theta_{mn}\in{\boldsymbol\Theta}, ~ \forall m\in[M], n\in[N],
\end{align}
where $\boldsymbol{\mathcal{H}}_R$ represents the set of channel vectors from RIS to all the receivers considered and $[M]$ is a shorthand for denoting the set $\{1,2,\dots,M\}$.
From \eqref{complex-sum}, we note that the composite channel gain of a user is essentially the absolute square of a summation of several complex numbers. Therefore, if we ignore the constraints, the maximum of the objective of \eqref{Prob} is attained only when the RIS reflecting phase $\theta_{mn}$ satisfies
\begin{equation}\label{eq-cond}
  \phi_{T, m}+\phi_{R, u, m}+\theta_{mn} = \varphi_u, ~ \forall m\in[M], \exists\boldsymbol{\psi}_n\in\boldsymbol{\mathcal{F}}, \forall u\in\mathcal{U},
\end{equation}
where $\varphi_u$ is an arbitrary constant phase value and $\mathcal{U}$ is the user index set. However, there are three defining factors making \eqref{eq-cond} inapplicable.
First, the values of $\phi_{T, m}, \phi_{R, u, m}, \forall m\in[M], \forall u\in\mathcal{U}$ are continuous, while $\theta_{mn}$ can only be selected from a discrete phase set $\boldsymbol\Theta$. This means that finding a common $\varphi_u$ that makes all the $\theta_{mn}, \forall m\in[M]$ valid is highly infeasible.
Second, the knowledge of $\phi_{T, m}, \phi_{R, u, m}, \forall m\in[M], \forall u\in\mathcal{U}$ is practically not available at the RIS.
Therefore, with the motivation of optimizing the codebook design of the RIS without explicit knowledge of the channels, we propose to leverage the powerful learning capability of reinforcement learning, which will be discussed in the next section.

\section{Deep Reinforcement Learning Based Reflection Codebook Design} \label{sec:Solution}

In this section, we discuss in detail the proposed approach for addressing problem \eqref{Prob}. As mentioned before, \eqref{eq-cond} is hard to solve mainly due to: (i) the lack of knowledge on $\phi_{T, m}, \phi_{R, u, m}$, (ii) finite size of reflection codebook, and (iii) discrete analog phase shifter constraints. Instead, we propose that the system only relies on quite limited information, such as the achieved received power at the receivers, to evaluate how good an reflection codebook performs and guide its optimization. In particular, we develop a deep reinforcement learning approach that explores the possible set of phases $\{\theta_{mn}, \forall m, n\}$ that approach the optimal receive power (the objective of \eqref{Prob}).
However, a simple calculation indicates that the number of possible reflection vectors increases exponentially with respect to the number of elements in RIS, with the base given by $2^q$. Besides, since the ultimate goal is a codebook, this further makes the searching space explode by its combination nature, which results in an intolerably slow convergence if all the phases are directly learned. To address this problem, we develop a fast convergent algorithm that quickly constructs an reflection codebook.
The proposed approach first clusters users into $N$ groups (the size of $\boldsymbol{\mathcal{F}}$). Then by going through a multi-level RIS sub-array design process, each group of users will be finally served by an optimized reflection vector. In the next subsections, we describe more details about the proposed reflection codebook learning framework.

\subsection{User Clustering}

Due to the finite size of the codebook, it is not possible to design user-specific reflection vector. Instead, the users sharing similar channels are served by the same reflection vector in the codebook.
Therefore, the first step of our proposed RIS reflection codebook design approach is to cluster the surrounding users based on the similarity of their channels. However, given that the explicit channel knowledge is not available, such clustering is performed based on a well-designed power-based feature matrix, where the power vector of each user is obtained by leveraging a set of sensing beams. The details of which could be found from Section V.A in our previous work \cite{Zhang2021Reinforcement}.

\textbf{Remark:} In practice, a newly deployed RIS might first use a random reflection codebook or a pre-defined codebook such as beamsteering codebook to serve the user. At the same time, it keeps accumulating such power vectors by listening to the reference signal received power (RSRP) feedback reported from the users during the beam training stage. Once it has enough beam training power vectors, the clustering can be performed to train a power-based user channel classifier for the reflection codebook learning purpose.

\subsection{Multi-Level RIS Codebook Design Architecture}

Even though we have decomposed the problem of learning an reflection codebook into $N$ independent and parallel sub-problems (for the $N$ user clusters), the task of learning a single reflection vector is still highly complex due to the large number of reflecting elements in RIS. Therefore, we develop a multi-level learning approach that aims to lower the computational complexity of each sub-problem.
In short, we first divide the whole surface into several sub-arrays (which could also represent distributed sub-surfaces) and learn the reflection vectors only for those sub-surfaces. In addition, we leverage the trained parameters of the DRL model for one of the sub-surface to further reduce the training time (iterations) needed for the other sub-surfaces (transfer learning of the initial neural network weights).
Then, we successively combine the learned reflection beamforming vectors of some sub-surfaces and refine the learning till we learn the full reflection vector of the full surface. Next, we describe in detail the basic idea of this multi-level learning process. For simplicity and clarity, we demonstrate our idea in single user case.
%
Therefore, we drop the indices for both reflection vectors in $\boldsymbol{\mathcal{F}}$ and users in $\mathcal{U}$.

\subsubsection{\textbf{Decompose the Large Surface}}

Without loss of generality, we assume that the whole array is divided into sub-arrays with equal size $P_1$ \footnote{It is worth noting that such decomposition of array could also mean distributed RIS, where each sub-array is a distributed surface}.
Further, we define $\alpha_m \triangleq \alpha_{T, m}\alpha_{R, m}$ and $\phi_m \triangleq \phi_{T, m}+\phi_{R, m}$. Then, the summation term in \eqref{complex-sum} can be written as
\begin{multline}\label{construct-level}
  \sum_{m=1}^M\alpha_me^{j(\phi_m+\theta_m)}=\sum_{p_v}^{P_v}e^{j\theta_{p_v}}\sum_{p_{v-1}}^{P_{v-1}}e^{j\theta_{p_v, p_{v-1}}}\dots \\
  \sum_{p_2}^{P_2}e^{j\theta_{p_v, \dots, p_2}}\sum_{p_1}^{P_1}\alpha_{p_v, \dots, p_1}e^{j(\phi_{p_v, \dots, p_1}+\theta_{p_v, \dots, p_1})},
\end{multline}
where we assume a $v$-level learning process and $M=P_vP_{v-1}\cdots P_1$. The conversion of the RIS element index $m$ to multi-level index $(p_v, p_{v-1}, \dots, p_1)$ for $\alpha_m$, $\phi_m$ and $\theta_m$ can be found as follows
\begin{equation}\label{index-conversion}
  p_{v^\prime}=\left\{\begin{array}{cc}
                        \tilde{p}_{v^\prime}, & \text{if $\tilde{p}_{v^\prime}\ne 0$}, \\
                        P_{v^\prime}, & \text{otherwise},
                      \end{array}\right. \forall v^\prime = 1,2,\dots,v,
\end{equation}
where
\begin{equation}\label{ind-conversion}
  \tilde{p}_{v^\prime} =~ \mathrm{mod}\left(\left\lceil \frac{m}{\prod_{k=1}^{v^\prime-1}P_k} \right\rceil, P_{v^\prime}\right),
\end{equation}
with $P_0=1$ assumed. Therefore, \eqref{construct-level} divides the original ``full-size'' reflection vector design problem into $v$ levels, with $v^\prime$ level a total number of $P_v\cdots P_{v^\prime}$ phases to be learned. However, the maximum number of phases that need to be learned \emph{simultaneously} throughout all the $v$ levels is only $\max\{P_1, P_2, \dots, P_v\}$. Such reduction in the number of phases significantly decreases the size of the searching space, making the algorithm converge fast, as will be shown in \sref{sec:NSch}. Besides, this multi-level framework also allows the system to configure the different  design process levels according to the number of RIS elements and its computation capability.

\subsubsection{\textbf{Sequential Design Process}}

The learning process starts from the lowest level, i.e., $v^\prime=1$, with $P_1$ phases to be learned ``simultaneously'' for each sub-array, which is only $1/(P_v\cdots P_2)$ of the original task.
The phases of level 2 and beyond, i.e. $v^\prime=2,\dots,v$, play a different role than the lowest level phases. They \textbf{combine} the lowest level phases somehow to form a higher dimensional beamforming vector that finally represents the full array.
Naively concatenating the lowest level phases together, however, will result in significant performance degradation due to the incoherent phases caused by the separate learning processes. Therefore, the similar learning process should also be conducted on specifying the high level combining phases. To illustrate this point, we consider a two-level architecture for simplicity. At the second level, the objective should be on aligning the $P_2$ sub-signals obtained at the first level. We assume that the $p_2$-th sub-signal is rotated by an angle of $\tilde{\theta}_{p_2}$, then the summation of those $P_2$ rotated sub-signals can be written as
\begin{equation}\label{combining}
  \sum_{p_2=1}^{P_2} e^{j\tilde{\theta}_{p_2}} \sum_{p_1=1}^{P_1}\alpha_{p_2, p_1}e^{j(\phi_{p_2, p_1}+\tilde{\theta}_{p_2, p_1})} = \sum_{p_2=1}^{P_2}\tilde{\alpha}_{p_2}e^{j(\tilde{\phi}_{p_2} + \tilde{\theta}_{p_2})},
\end{equation}
where we ignore the normalization factor for the reflection vector, and $\{\tilde{\theta}_{p_2, p_1}, \forall p_2=1,\dots,P_2, p_1=1,\dots,P_1\}$ are the learned first level phases.
Therefore, for the $p_2$-th sub-array, the \textbf{effective} phase shift executed by the $p_1$-th element is $\tilde{\theta}_{p_2, p_1}+\tilde{\theta}_{p_2}$. To uphold the discrete phase shifter constraint, both $\tilde{\theta}_{p_2, p_1}$ and $\tilde{\theta}_{p_2}$ are selected from $\boldsymbol\Theta$, and hence $\tilde{\theta}_{p_2, p_1}+\tilde{\theta}_{p_2}\in\boldsymbol\Theta$.
In general, the ultimate phase for the $m$-th element in the RIS can be expressed as
\begin{equation}\label{final-ph}
  \tilde{\theta}_m = \tilde{\theta}_{p_v, \dots, p_1} + \tilde{\theta}_{p_v, \dots, p_2} + \cdots + \tilde{\theta}_{p_v},
\end{equation}
where the index conversion is given by \eqref{index-conversion}. Similarly, since all the phases at the right hand side of \eqref{final-ph} are selected from $\boldsymbol\Theta$, the synthesized $\tilde{\theta}_m$ is also a valid phase value.

\subsection{Reinforcement Learning Formulation}

The phase design process at each level is conducted in a reinforcement learning fashion. As the problem features a finite yet very huge action space, we propose using a novel architecture called Wolpertinger \cite{Dulacarnold2015} to efficiently explore optimal policy in a large discrete action space.
Furthermore, we specify the indispensable components of reinforcement learning in the context of the phase design problem as follows:
\begin{itemize}
  \item \textbf{State:} We define the state ${\bf s}_t$ as a vector consisting of all the phase values at the $t$-th iteration, that is, ${\bf s}_t=\left[\theta_1, \theta_2, \dots, \theta_{M^\prime}\right]^T$, with $M^\prime=P_{v^\prime}$ at the $v^\prime$ level.
  \item \textbf{Action:} We define the action ${\bf a}_t$ as the element-wise changes to all the phases in ${\bf s}_t$. Since the phases can only take values in $\boldsymbol\Theta$, a change of a phase means that the phase shifter selects a value from $\boldsymbol\Theta$. Therefore, the action is also directly specified as the next state, i.e., ${\bf s}_{t+1}={\bf a}_t$. 
  \item \textbf{Reward:} We define a binary reward mechanism, i.e. the reward $r_t$ takes values from $\{+1, -1\}$. We compare the composite channel gain achieved by the current reflection vector, denoted by $g_t$, with the previous gain $g_{t-1}$. If $g_t > g_{t-1}$, then $r_t=+1$. Otherwise, $r_t=-1$.
\end{itemize}

It is worth noting that such learning framework does not rely on the channel state information. The DRL agent is capable of adjusting its decision for choosing phases solely based on the receivers' feedbacks.

\begin{figure*}[t]
	\centering
	\subfigure[RIS with stationary channels]{ \includegraphics[width=0.34\linewidth]{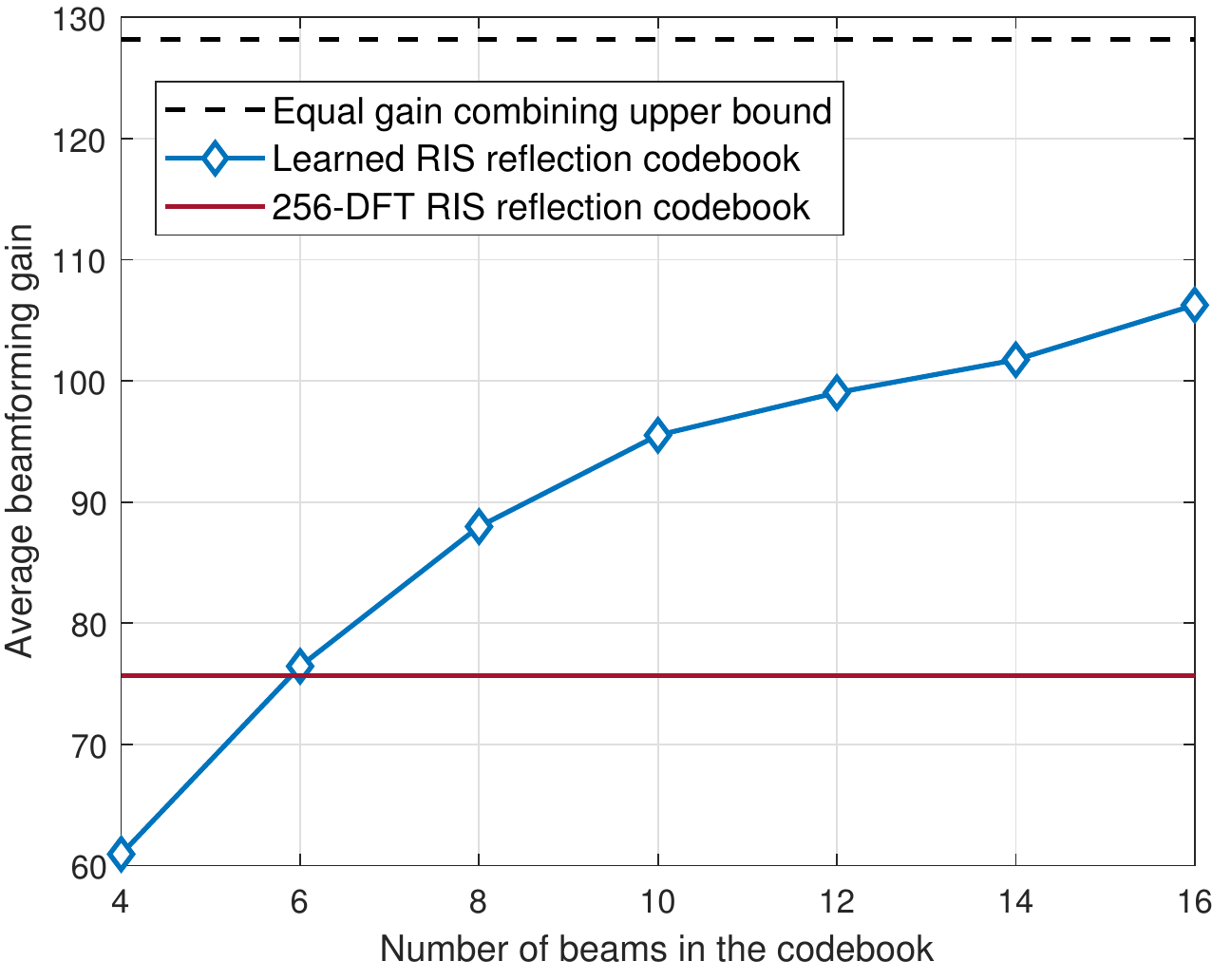} \label{pf-vs-beams} }
	\subfigure[Distributed RISs with non-stationary channels]{ \includegraphics[width=0.61\linewidth]{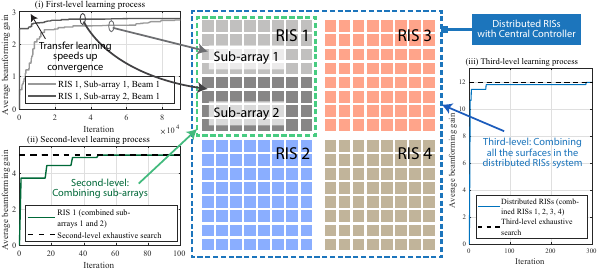} \label{pf-vs-iters} }
	\caption{The learning results of the proposed DRL-based multi-level RIS reflection codebook learning solution on: (a) the stationary channels, where the average beamforming gain versus the number of beams in the codebook is depicted, and (b) the non-stationary RIS, where the average beamforming gain (achieved by a single reflection vector on a single user cluster) versus number of iterations at different learning levels are depicted.}
	\label{ddd}
\end{figure*}

\section{Simulation Results} \label{sec:Simu}

In this section, we evaluate the performance of our proposed DRL based multi-level learning approach for RIS reflection codebook design. We first describe the adopted scenario and dataset used in our simulations and then discuss the results.

\subsection{Scenario and Dataset}

In our simulations, we consider the outdoor scenario `O1\_60' which is offered by the DeepMIMO dataset \cite{DeepMIMO}.
We first generate the channels between every receiver $u$ to the RIS, i.e., $\boldsymbol{\mathcal{H}}_R$, and the channel between a transmitter to the RIS, i.e., $\bh_T$. Then, we generate the composite channels $\bh_T\odot\bh_{R, u}, \forall\bh_{R, u}\in\boldsymbol{\mathcal{H}}_R$.
We adopt the following DeepMIMO parameters: (1) Scenario name: O1\_60, (2) Active BSs: 3, (3) Active users: Row 1201 to 1400, (4) Number of BS antennas in (x, y, z): (1, 256, 1), (5) System bandwidth: 1 GHz, (6) Number of multipaths: 5.
We further select 80 out of 181 users each row, yielding a total number of 16,000 users. The transmitter is at row 850 and column 90 in the `O1\_60' scenario.

We also generate another dataset that takes non-stationarity into consideration. We consider a scenario where there are 4 \textit{geographically distributed} reflecting surfaces, and each one of them adopts a 64-element uniform linear array. These surfaces maintain a distance of 1 meter between each other spatially, and are aligned along the y-axis in the `O1\_60' scenario. We assume that this distributed RIS system is serving the same grid of users as the previous dataset, and the transmitter is at the same position. The channels between the users and each surface are generated independently to account for the non-stationarity effect caused by the huge aperture of the distributed RIS system. Finally, the channels for different surfaces are concatenated together to form the ultimate channels from the users to this distributed RIS surface.

\subsection{Evaluation on Stationary Channel}

We first consider a stationary RIS with 256 reflecting elements, where each reflecting element is equipped with a 4-bit quantized phase shifter. Besides, we adopt a two-level learning model with $P_1=32$ and $P_2=8$. In other words, we first divide this 256-element RIS into 8 sub-arrays, with each one of the sub-arrays having 32 elements. During the first stage, 8 \textit{independent} reflection vectors are learned for those sub-arrays. To be more specific, we first train a model to learn one reflection vector for one of the sub-arrays, and then use the parameters of the trained deep learning model to initialize the other 7 models to reduce the convergence time. And at the second stage, 8 combining phases are learned to \textit{coherently} synthesize the ultimate reflection vector.
\fref{pf-vs-beams} plots the average beamforming gain achieved by the proposed approach versus the size of the codebook. As can be seen in the figure, \textbf{the developed approach is able to learn a 16-beam codebook that brings more than $40\%$ improvement on the average beamforming gain performance over a 256-DFT codebook, while requiring only $6.25\%$ of the beam training overhead.}

\subsection{Evaluation on Non-stationary Channel} \label{sec:NSch}

Next, we consider a distributed RIS-assisted wireless system. As explained before, due to the distinct physical separations, the channels across different surfaces may appear different patterns. This causes the distributed RISs, if viewed as a whole array, non-stationary.
Therefore, we adopt a three-level learning model with $P_1=32$, $P_2=2$ and $P_3=4$, that is, we first learn reflection vectors for each surface at the first and second stages, and then combine them at the third stage. Within each RIS, we further divide the 64-element surface into two 32-element sub-arrays to speed up the learning process, as used before.
\fref{pf-vs-iters} shows the achieved average beamforming gain versus the number of iterations for different learning levels. \fref{pf-vs-iters} (i) illustrates the learning process of the first level, i.e. learning reflection vectors for sub-arrays of each surface. As can be seen, by initializing the parameters of the learning model for sub-array 2 with the well-trained sub-array 1's, with negligible number of iterations, the reflection vector for sub-array 2 can be learned. In \fref{pf-vs-iters} (ii), we plot the second-level learning process, the result of which forms an reflection vector for the first RIS.
Finally, 4 third-level combining phases are learned to combine the ultimate reflection vector for the distributed RIS system. As shown in \fref{pf-vs-iters} (iii), with less than 100 iterations, the model can achieve nearly the same performance as the third-level exhaustive search, which requires $(2^4)^4=65536$ iterations.

\section{Conclusion}

In this letter, we developed a DRL based learning framework for constructing the  reflection beam codebooks for large RIS surfaces. The developed solution adopts a multi-level learning framework that speeds up the convergence, and reduces the computational complexity. Simulation results highlight the capability of the proposed solution to efficiently learn beam codebooks that adapt to user distributions and channel characteristics, even for distributed RISs and non-stationary channels. Further, the results show the significant improvement over DFT codebooks in reducing the required codebook size.

\end{document}